\documentclass{article}
\usepackage[english]{babel}
\usepackage{amsmath,amssymb,tabularx,bm,array,booktabs,authblk}
\usepackage{xcolor,slashed,epsfig,cite,float}
\usepackage{verbatim,slashed,ulem,geometry,setspace,algorithm}
\usepackage{algorithmic,mathrsfs,braket,multirow,dcolumn,subfigure,caption}
\usepackage{graphicx}
\usepackage{threeparttable}
\graphicspath{{./Figures/}}
\usepackage[colorlinks=true,citecolor=blue,urlcolor=cyan,filecolor=magenta,linkcolor=red,frenchlinks=true,bookmarks]{hyperref}
\allowdisplaybreaks[4]
\renewcommand\sout{\bgroup \color{red}  \ULdepth=-.5ex \ULset}

\begin{document}

\title{A systematical study of the nucleon form factors with the pion cloud effect}

\newcommand{\email}[1]{\thanks{\href{mailto:#1}{#1}}}

\author[a,b]{Jiaqi Wang\email{jqwang@ihep.ac.cn}}
\author[c]{Dongyan Fu\email{fudongyan@impcas.ac.cn}}
\author[a,b]{Yubing Dong\email{dongyb@ihep.ac.cn}}

\affil[a]{Institute of High Energy Physics, Chinese Academy of Sciences, 
Beijing 100049, China}
\affil[b]{School of Physical Sciences, University of Chinese Academy 
of Sciences, Beijing 101408, China}
\affil[c]{Southern Center for Nuclear-Science Theory (SCNT),
\authorcr
Institute of Modern Physics, Chinese Academy of Sciences, Huizhou 516000, China}

\maketitle
\begin{abstract}
    The electromagnetic and gravitational form factors of the nucleon are studied simultaneously using a covariant quark-diquark approach, and the pion cloud effect on the form factors is explicitly discussed. In this study, the electromagnetic form factors are calculated to determine the parameters of our approach. Then, the gravitational form factors of the nucleon are evaluated with the same parameters. The electromagnetic and mechanical properties, such as charge radii, magnetic moments, and mass radii are presented. Moreover, the D-term, which is believed to be connected with the internal force inside the nucleon, is also discussed in this paper. Our results are in reasonable agreement with the experimental and lattice results, and it is suggested that the pion cloud contribution plays an indispensable role, especially for the magnetic moments and the sign of D-term.
\end{abstract}

\section{Introduction}

    Understanding the particle structure remains a fundamental topic in hadron physics. Form factors carry a wealth of information, serving as a useful tool for exploring the internal structure of particles. It is well known that for a spin-1/2 hadron, its electromagnetic form factors (EMFFs) are related to the electromagnetic properties such as the charge radius and magnetic moment. Similarly, the gravitational form factors (GFFs), which are derived from the matrix element of the symmetric energy-momentum tensor (EMT)~\cite{Pagels:1966zza}, could provide information about the interior of the particle such as the mass and spin distributions. The D-term extracted from the EMT matrix elements $T^{ij}$ is known as the ``last unknown global property" of the particle. It is believed that the D-term is related to the spatial deformations inside the particle and characterizes the distributions of the pressure and shear force \cite{Polyakov:2018zvc}.

    A significant number of experiments have been carried out to measure the EMFFs of the nucleon~\cite{Punjabi:2015bba}. However, due to the impracticality of directly detecting the coupling of the graviton with the matter field, the GFFs cannot be experimentally assessed. Fortunately, with the help of the deeply virtual Compton scattering (DVCS)\cite{Ji:1998pc}, a process in which the parton within the particle absorb a virtual photon and emit a real photon, one can extract the GFFs from the generalized parton distributions via the sum rules~\cite{Polyakov:2002yz,Diehl:2003ny}. In 2018, the quark contribution to the proton GFF $D_0(t)$ was first extracted, revealing the pressure~\cite{Burkert:2018bqq} and shear force~\cite{Burkert:2021ith} contributed by the quark. Subsequently, in 2024, Ref.~\cite{Cao:2024zlf} presented the first model-independent precise determinations of the GFFs. Despite the experimental challenges in detecting the GFFs, they can be investigated theoretically through the process analogous to that used for the EMFFs. The GFFs of the nucleon have been studied with different models including the lattice QCD~\cite{Mathur:1999uf,LHPC:2007blg,Hackett:2023rif}, the chiral quark-soliton model~\cite{Goeke:2007fp,Kim:2021jjf}, the Skyrme model~\cite{Kim:2012ts}, the light-front quark-diquark model~\cite{Chakrabarti:2020kdc,Choudhary:2022den}, the large $N_c$ quark model~\cite{Lorce:2022cle}, the light-cone QCD sum rules~\cite{Azizi:2019ytx,Anikin:2019kwi}, the $\pi-\rho-\omega$ soliton model~\cite{Jung:2014jja}, and so on~\cite{Burkert:2023wzr}.  

    In this paper, the form factors of the nucleon are calculated with a covariant quark-diquark approach, which has been applied to our previous studies on the decuplet baryons~\cite{Wang:2023bjp,Fu:2022rkn}. The nucleon, a three-body system, is treated as the combination of a quark and a diquark, which could effectively simplify the numerical calculations. Furthermore, the dressed quark with the pion cloud instead of the point-like quark is employed in this study. 
         
    In some works focusing on the EMFFs of the nucleon and other hardons~\cite{Thomas:1981vc,Cloet:2014rja,Fu:2025vkq}, it has been demonstrated that the pion cloud plays an important role in the form factors. Especially for the nucleon, it results in a notable enhancement in the magnitude of the magnetic moments and influences the shape of the electric form factor of the neutron obviously. However, the effect of the pion cloud on the GFFs has been rarely discussed. In this paper, the EMFFs and GFFs of the nucleon are calculated with the same parameters simultaneously to evaluate the influence of the pion cloud. 
    
    This paper is organized as follows. Section II gives the form factor definitions of the spin-1/2 particle and introduces the covariant quark-diquark approach with the pion cloud. In Sec. III, the numerical results of the form factors are given, and the electromagnetic and mechanical properties of the nucleon are obtained. Furthermore, the effect of the pion cloud on the EMFFs and the GFFs is discussed in detail. Finally, a brief discussion and a summary are displayed in Sec. IV.

\section{Form Factors And Quark-diquark Approach}

\subsection{Form factors of the nucleon}

For the nucleon, a spin-1/2 particle, the matrix element of the electromagnetic current $J^\mu$ is expressed as~\cite{Cotogno:2019vjb}
\begin{equation}
\begin{split}
\label{JMatrix}
    \langle p',\lambda'|\hat{J}^{\mu}|p,\lambda\rangle&=\overline{u}\left(p',\lambda'\right)\biggl[\gamma^\mu F_{1}\left(t\right)+\frac{i \sigma^{\mu q}}{2M_N}F_{2}\left(t\right)\biggr]u\left(p,\lambda\right),
\end{split}
\end{equation}	
where $i \sigma^{\mu q}=i \sigma^{\mu\rho}q_\rho$, $M_N$ stands for the mass of the nucleon, and $u\left(p,\lambda\right)$ is the Dirac spinor with normalization as $\overline{u}\left(p,\lambda\right)u\left(p,\lambda\right)=2M_N$. The kinematical variable $q$ is defined as $q^\mu=p'^\mu-p^\mu$ and $t=q^2$, where $p$ ($p'$) is the initial (final) momentum of the nucleon. 

The electric and magnetic form factors are defined as~\cite{Ernst:1960zza}
\begin{subequations}
\label{EMFFs}
\begin{align}
    &G_E(t)=F_1(t)+\frac{t}{4M_N^2} F_2(t),
\\
    &G_M(t)=F_1(t)+F_2(t).
\end{align}
\end{subequations} 
When the squared transfer momentum $t$ goes to zero, one can obtain the charge $Q_e=G_E(0)$ and the magnetic moment $\mu=\frac{e}{2M_N}G_M(0)$.
Moreover, the charge and magnetic radii of the particle can be derived from their form factors,
\begin{equation}
\label{EMRadius}
    {\langle r^2_{E}\rangle}=\left.\frac{6}{G_{E}(0)} \frac{d}{dt}G_{E}(t)\right|_{t=0}, 
    \quad {\langle r^2_{M}\rangle}=\left.\frac{6}{G_{M}(0)} \frac{d}{dt}G_{M}(t)\right|_{t=0}.    
\end{equation}
For the neutral neutron, the charge radius has the special definition,
\begin{equation}
\label{NERadius}
{\langle r^2_{E}\rangle}=\left.6 \frac{d}{dt}G_{E}(t)\right|_{t=0}.
\end{equation}

The GFFs of the nucleon are extracted from the matrix element of the EMT current~\cite{Cotogno:2019vjb},
\begin{equation}
\begin{split}
\label{TMatrix}
    \langle p',\lambda'|\hat{T}^{\mu \nu}|p,\lambda\rangle&=\overline{u}\left(p',\lambda'\right)\biggl[\frac{ P^\mu P^\nu}{M_N} F_{1,0}(t)+\frac{\left(q^\mu q^\nu-g^{\mu\nu}q^2\right)}{M_N}F_{2,0}(t)+M_N g^{\mu\nu}F_{3,0}(t)\\
    &+\frac{i}{4M_N}P^{ \{ \mu}\sigma^{\nu\}q}F_{4,0}(t)\biggl] u(p,\lambda),
\end{split}
\end{equation}
where $P^{ \{ \mu}\sigma^{\nu\}q}=P^\mu\sigma^{\nu q}+P^\nu\sigma^{\mu q}$ and $P^\mu=\left(p'^\mu+p^\mu\right)/2$. Notice that $F_{3,0}(t)$ is a non-conserving term which will vanish when considering the total EMT of the quark and the gluon. Thus, we will simply ignore it in the results.

In the Breit frame, $p=(E,-\bm{q}/2)$ and $p'=(E,\bm{q}/2)$, the gravitational multipole form factors (GMFFs), $\varepsilon_0(t)$, $\mathcal{J}_1(t)$ and $D_0(t)$, are the combinations of the GFFs $F_{i,0}(t)\, (i=1,2,3,4)$~\cite{Goeke:2007fp},
\begin{subequations}
\label{GFFs}
\begin{align}
    \varepsilon_0(t)=F_{1,0}(t)+&\frac{t}{4M_N^2}\left[-F_{1,0}(t)-4 F_{2,0}(t)+F_{4,0}(t)\right],
\\
    &\mathcal{J}_1(t)=\frac{1}{2}F_{4,0}(t),
\\
    &D_0(t)=4F_{2,0}(t).
\end{align}
\end{subequations} 
\noindent
In the following, we will uniformly use GFFs, without distinguishing between the GFFs and the GMFFs, and one can discriminate them according to the context.
$\varepsilon_0(t)$ and $\mathcal{J}_1(t)$ are related to the mass and the spin distributions of the particle, respectively. The mass radius of the particle can be obtained from $\varepsilon_0(t)$, 
\begin{equation}
\label{MassRadius}
    \langle r^2_m \rangle = \left.\frac{6}{\varepsilon_0(0)} \frac{d}{dt}  \varepsilon_0(t) \right|_{t=0}.
\end{equation}

By applying the Fourier transformation to $\varepsilon_0(t)$ in the coordinate space, one can get the energy density of the particle,
\begin{equation}
\label{EnergyDensity}
    \mathcal{E}_0(r)=M_N \int \frac{d^3 q}{(2\pi)^3}e^{-i\bm{q}\cdot \bm{r}}\varepsilon_0(t).
\end{equation}
Moreover, the angular momentum density is defined as~\cite{Lorce:2017wkb,Schweitzer:2019kkd}
\begin{equation}
\begin{split}
\label{AMDensity}  
    J^i(\bm{r},\bm{s})&=J^i_{mono}(\bm{r},\bm{s})+J^i_{quad}(\bm{r},\bm{s})\\
    &=s^i\int \frac{d^3 q}{(2\pi)^3} e^{-i\bm{q}\cdot \bm{r}}\left(\mathcal{J}_1(t)+\frac{2}{3}t \frac{d \mathcal{J}_1(t)}{d t}\right)+s^j\int \frac{d^3 q}{(2\pi)^3} e^{-i\bm{q}\cdot \bm{r}}\left(q^i q^j - \frac{1}{3}\bm{q}^2\delta^{ij}\right)\frac{d \mathcal{J}_1(t)}{d t},
\end{split}
\end{equation}
where the first and second terms represent the monopole and the quadrupole contributions, respectively. Meanwhile, the monopole and quadrupole densities can be expressed as~\cite{Schweitzer:2019kkd}
\begin{equation}
\label{MQDensity}
    J^i_{mono}(\bm{r},\bm{s})=s^i\rho_J(r), \quad J^i_{quad}(\bm{r},\bm{s})=s^j(e_r^ie_r^j - \frac{1}{3}\delta^{ij})b(r),
\end{equation}
where $e_r^i=r^i/r$. The spin density $\rho_J(r)$ connects $J^i_{mono}$ and $J^i_{quad}$ through the relation,
\begin{equation}
\label{MQRelation}
    b(r)=-\frac{3}{2}\rho_J(r),
\end{equation}
and has the property $\int d^3 r \rho_J(r)=\frac{1}{2}$.      

$D_0(t)$, related with the so-called D-term $D=D_0(0)$, is believed to be connected with the pressure and shear force distributions inside the particle in the classical physics, which can be expressed as\cite{Polyakov:2018zvc}
\begin{equation}
\label{Pressure}
    p(r)=\frac{1}{6M_N}\frac{1}{r^2}\frac{d}{dr}r^2\frac{d}{dr}\widetilde{D}_0(r),
\end{equation}
\begin{equation}
\label{ShearForce}
    s(r)=-\frac{1}{4M_N}r\frac{d}{dr}\frac{1}{r}\frac{d}{dr}\widetilde{D}_0(r),
\end{equation}
where $\widetilde{D}_0(r)=\int \frac{d^3 q}{(2\pi)^3}e^{-i\bm{q}\cdot \bm{r}}D_0(t)$. The pressure and shear force satisfy the equilibrium relation,
\begin{equation}
    \frac{2}{3}\frac{ds(r)}{dr}+2\frac{s(r)}{r}+\frac{dp(r)}{dr}=0.
\end{equation}
$\frac{2}{3}s(r)+p(r)$ characterizes the normal force distribution in the hardon. Similarly, the mechanical radius \cite{Polyakov:2018zvc} can be derived from the combination as
\begin{equation}
\label{MechRadius}
    \langle r^2_\text{mech} \rangle = \frac{\int d^3 r r^2 \left[\frac{2}{3}s(r)+p(r)\right]}{\int d^3 r \left[\frac{2}{3}s(r)+p(r)\right]}=\frac{6D}{\int^0_{-\infty}dt D_0(t)}.
\end{equation}

\subsection{Quark-diquark approach}\label{QDApproach}

To simplify the calculation, the nucleon is treated as the combination of a quark and either a scalar (spin-0) diquark or an axialvector (spin-1) diquark. The flavor wave functions of the proton and the neutron are written as~\cite{Ma:2002ir}
\begin{subequations}
\label{WaveFunc}
\begin{align}
    &|p\rangle=\cos{\theta}|u(ud)_s\rangle+\sin{\theta}\left[\sqrt{\frac{2}{3}}|d (uu)_a\rangle-\sqrt{\frac{1}{3}}|u (ud)_a\rangle\right],\\
    &|n\rangle=\cos{\theta}|d(ud)_s\rangle+\sin{\theta}\left[-\sqrt{\frac{2}{3}}|u (dd)_a\rangle+\sqrt{\frac{1}{3}}|d(ud)_a\rangle\right],
\end{align}
\end{subequations} 
where $(q_1q_2)_s$ and $(q_1q_2)_a$ respectively represent the scalar and the axialvector diquarks consisting of quarks $q_1$ and $q_2$, and $\theta$ is the mixing angle between the scalar and the axialvector diquarks.

\begin{figure}[h]
    \centering
    \subfigure[]{\includegraphics[scale=0.65]{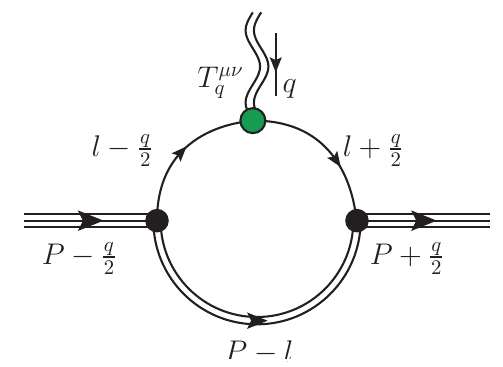}\label{tq}}
    \subfigure[]{\includegraphics[scale=0.65]{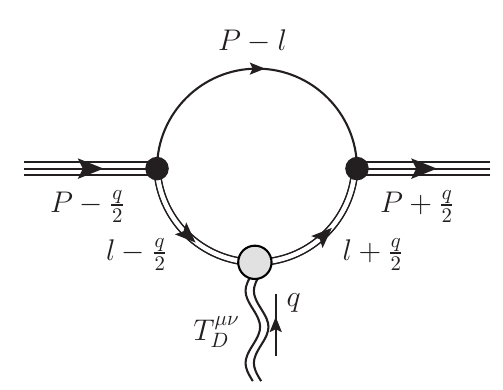}\label{td}}
    \subfigure[]{\includegraphics[scale=0.65]{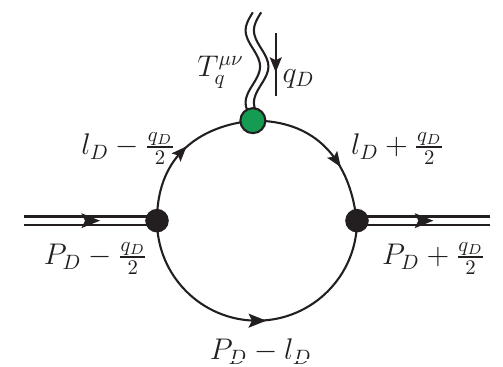}\label{td0}}
    \caption{\small{Feynman diagrams for the coupling processes of the EMT with the quark (a) and the diquark (b). When the EMT current couples with the diquark, the internal structure of the diquark (c) is considered as well. The green points stand for the EMT vertex of the dressed quark introduced in the next subsection.}}
\label{EMT-Nucleon}
\end{figure}

The GFFs of the nucleon are derived from the matrix element of the EMT, which can be written as the sum of the contributions from the quark and the diquark,
\begin{equation}\label{eq-emt}
    \langle p',\lambda'|\hat{T}^{\mu\nu}|p,\lambda\rangle=\langle p',\lambda'|\hat{T}^{\mu\nu}_q|p,\lambda\rangle+\langle p',\lambda'|\hat{T}^{\mu\nu}_D|p,\lambda\rangle.
\end{equation}
Corresponding to Eq.~\eqref{eq-emt},
Figs.~\ref{tq} and \ref{td} show the Feynman diagrams of the coupling processes with the quark and the diquark, respectively.
When calculating the EMFFs, one merely needs to replace the EMT current $T^{\mu\nu}$ with the electromagnetic current $J^\mu$. The detailed computation procedure is presented in our previous work on $\Delta$ resonance~\cite{Fu:2022rkn}, and here we provide a concise overview of the GFF calculation framework.

Since there are two kinds of the diquark in Fig.~\ref{EMT-Nucleon}, the scalar or the axialvector one, the form factors should be discussed separately. The quark contribution of the matrix element has two parts corresponding to the diquark,
\begin{subequations}
\label{QuarkT}
\begin{align}
    \langle p',\lambda'|\hat{T}^{\mu\nu}_{q(s)}|p,\lambda\rangle=&\overline{u}(p',\lambda')\int\frac{d^4l}{(2\pi)^4}\frac{-i c^2}{\mathcal{D}}\Gamma_s (\slashed{l}+\frac{\slashed{q} }{2}+m_q)\Lambda^{\mu\nu}(t)(\slashed{l}-\frac{\slashed{q} }{2}+m_q)\Gamma_s u(p,\lambda),\label{SQuarkT}
\\
\begin{split}
    \langle p',\lambda'|\hat{T}^{\mu\nu}_{q(a)}|p,\lambda\rangle=&\overline{u}(p',\lambda')\int\frac{d^4l}{(2\pi)^4}\frac{-i c^2}{\mathcal{D}}\Gamma_a^{\beta'} (\slashed{l}+\frac{\slashed{q} }{2}+m_q)\\
    &\times \Lambda^{\mu\nu}(t)(\slashed{l}-\frac{\slashed{q} }{2}+m_q)\left[-g_{\beta \beta'}+\frac{(P_\beta-l_\beta)( P_{\beta'}-l_{\beta'})}{m_D^2}\right]\Gamma_a^{\beta} u(p,\lambda),\label{AQuarkT} 
\end{split}
\end{align}
\end{subequations} 
where the subscript $q$ stands for the quark contribution and $s$ ($a$) means that the quark is from the nucleon with the scalar (axialvector) diquark.
$\Gamma_s$ ($\Gamma_a^\mu$), $\mathcal{D}$ and $\Lambda^{\mu\nu}(t)$ respectively stand for the quark-diquark-nucleon vertex for the scalar (axialvector) diquark, all the denominators, and the interaction vertex between the quark and the EMT current, which will be discussed in Sec. \ref{PionCloud}.

Specially note that in our previous paper about $\Delta(1232)$, we neglected the momentum part (the second term) of the axialvector diquark propagator in order to have finite results \cite{Fu:2022rkn}. In this study, the momentum part does not lead to the divergence. As a consequence, this term is considered during our calculation. In the obtained results, this term only negatively contributes to the magnetic moments of the proton and neutron about 3\% and 1\%. Therefore, it is reasonable to neglect this term in our previous study.

$\Gamma_s$ and $\Gamma_a^\beta$ in Eq.(18) stand for the quark-diquark-nucleon vertex borrowed from Ref.~\cite{Scadron:1968zz} and respectively read
\begin{equation}
\label{Vertex}
    \Gamma_s=c_s,\quad \Gamma_a^\mu=c_a(\gamma^\mu+g_1 \frac{p_r^\mu}{M_N})\gamma^5,
\end{equation}
where $p_r$ stands for the relative momentum between the quark and the diquark and $g_1$ is a coupling constant to be determined.

An extra scalar function $\Xi(p_1,p_2)$ is attached to the vertex to simulate the bound state of the nucleon. Generally, the scalar function should be obtained through solving the Bethe-Salpeter equation (BSE). Here, we take an ansatz to simplify the calculation~\cite{Frederico:2009fk},
\begin{equation}
\label{sfunc}
    \Xi(p_1,p_2)=\frac{c}{(p_1^2-m_R^2+i\epsilon)(p_2^2-m_R^2+i\epsilon)},
\end{equation}
where $p_1$ and $p_2$ stand for the momenta of the quark and the diquark, $c$ is a coefficient in units of $\text{GeV}^4$, and $m_R$ in Eq.~\eqref{sfunc} is a cut-off parameter to be determined. The factor $\mathcal{D}$ in Eq.~\eqref{QuarkT}  stands for all the denominators from the propagators and the scalar functions as
\begin{equation}
\begin{split}
\label{Propagator}
    \mathcal{D}=&\left[\left(l+\frac{q}{2}\right)^2-m_q^2+i\epsilon\right]\left[\left(l-\frac{q}{2}\right)^2-m_q^2+i\epsilon\right]\left[\left(l-P\right)^2-m_D^2+i\epsilon\right]\\
    &\times\left[\left(l+\frac{q}{2}\right)^2-m_R^2+i\epsilon\right]\left[\left(l-\frac{q}{2}\right)^2-m_R^2+i\epsilon\right]\left[\left(l-P\right)^2-m_R^2+i\epsilon\right]^2.
\end{split}
\end{equation}	

The matrix element contributed by the diquark can also be evaluated in the similar way. For the scalar diquark, the matrix element reads\footnote{In our previous work on $\Delta$ resonance~\cite{Fu:2022rkn}, an extra quark propagator is added in the expression of the scalar diquark contribution (Eq. (33)). In Eq. \eqref{SDiquarkT}, it is corrected.}

\begin{equation}
\begin{split}
\label{SDiquarkT}
   \langle p',\lambda'|\hat{T}^{\mu\nu}_{D(s)}|p,\lambda\rangle&=\overline{u}(p',\lambda')\int\frac{d^4l}{(2\pi)^4}\frac{i c^2}{\mathcal{D}'}\Gamma_s (\slashed{P}-\slashed{l}+m_q)X^{\mu\nu}\Gamma_s u(p,\lambda).
\end{split}
\end{equation}	
The structure of $\mathcal{D}'$ is similar with Eq.~\eqref{Propagator} with only $m_q$ and $m_D$ exchanged. $X^{\mu\nu}$ is the interaction of the spin-0 diquark EMT current, which is derived from the inner structure of the diquark originated from Fig.~\ref{td0} as
\begin{equation}
    X^{\mu\nu}=\sum_q\langle p_D' ,\lambda_D'|\hat{T}^{\mu\nu}_{q}|p_D,\lambda_D\rangle.
\label{SDiquarkStructure}
\end{equation}  
The effective current can be expressed as~\cite{Cotogno:2019vjb}
\begin{equation}
   X^{\mu\nu}=2 P^\mu_D P^\nu_D F_{1,0}^D(t)+2\left(q^\mu q^\nu-g^{\mu\nu}q^2\right)F_{2,0}^D(t)+2m_D^2 g^{\mu\nu} F_{3,0}^D(t),
\label{SEffectiveEMTs}
\end{equation}    
where $P_D=(p'_D+p_D)/2$, $q=q_D=p'_D - p_D$ with $p_D$ ($p'_D$) as the initial (final) momentum of the diquark, and $F_{1,0}^D(t), F_{2,0}^D(t)$ and $F_{3,0}^D(t)$ are the form factors of the scalar diquark obtained with the similar procedure in Eqs.~(\ref{QuarkT}-\ref{Propagator}). The quark-quark-diquark vertex is borrowed from Ref.~\cite{Meyer:1994cn}, which reads $\gamma^5$ for the scalar diquark and $\gamma^{\mu}$ for the axialvector diquark.

For the axialvector diquark, the effective current $X^{\mu\nu,\beta' \beta}$ can be expressed as~\cite{Cotogno:2019vjb}
\begin{equation}
\begin{split}
    &-\epsilon^*_{\beta'}(p'_D,\lambda'_D) X^{\mu\nu,\beta' \beta}\epsilon_{\beta}(p_D,\lambda_D)=\sum_q\langle p_D',\lambda'_D|\hat{T}^{\mu\nu}_{q}|p_D,\lambda_D\rangle\\
    &=-\epsilon^*_{\beta'}(p'_D,\lambda'_D)\biggl[ \frac{P^\mu_D P^\nu_D}{m_D} \left(g^{\beta' \beta}F_{1,0}^D(t)-\frac{q^{\beta'} q^{\beta}}{2m_D^2}F_{1,1}^D(t) \right)\\
    &+\frac{\left(q^\mu q^\nu-g^{\mu\nu}q^2\right)}{m_D}\left(g^{\beta' \beta}F_{2,0}^D(t)-\frac{q^{\beta'} q^{\beta}}{2m_D^2}F_{2,1}^D(t) \right)
    +m_D g^{\mu\nu}\left(g^{\beta' \beta}F_{3,0}^D(t)-\frac{q^{\beta'} q^{\beta}}{2m_D^2}F_{3,1}^D(t) \right)  \\
    &-\frac{P_D^{ \{ \mu}g^{\nu\}[\beta'}q^{\beta]}}{2m_D}F_{4,0}^D(t)-\frac{1}{2m_D}\left(q^{ \{ \mu}g^{\nu\}\{\beta'}q^{\beta\}}-2q^{\beta'} q^{\beta}g^{\mu\nu}-g^{\beta' \{ \mu}g^{\nu\}\beta}q^2\right)F_{5,0}^D(t)\\
    &+\frac{1}{2}m_D g^{\beta'\{\mu}g^{\nu\}\beta}F_{6,0}^D(t)-\frac{1}{2m_D}P^{[\mu}g^{\nu][\alpha'}\Delta^{\alpha]}F_{8,0}^D(t)-\frac{1}{2m_D}\Delta^{[\mu}g^{\nu][\alpha'}\Delta^{\alpha]}F_{9,0}^D(t)\biggl]\epsilon_{\beta}(p_D,\lambda_D).
\end{split}
\label{ADiquarkStructure}
\end{equation} 

The transition current between the scalar diquark and the axialvector diquark is also taken into consideration in this work. The effective currents of the diquark internal EMFFs and GFFs are written as

\begin{subequations}
\label{TranCurrent}
\begin{align}
    j^{\mu,\alpha}_{s\rightarrow a}&=-j^{\mu,\alpha}_{a\rightarrow s}=i \epsilon^{\mu \alpha  P \Delta}F_{sa}(t),\\
    X^{\mu\nu,\alpha}_{s\rightarrow a}=&-X^{\mu\nu,\alpha}_{a\rightarrow s}=i P^{\{\mu} \epsilon^{\nu\}\alpha P \Delta} F_{sa}(t),
\end{align}
\end{subequations} 
where the subscripts $s\rightarrow a$ and $a\rightarrow s$ present the transition process from the scalar diquark to the axialvector diquark and its reverse process. For the EMFFs, only the momentum-related term in the axial-vector vertex $\Gamma^\mu_a$ (seen in Eq. \eqref{Vertex}) contributes to the transition part, while for the GFFs, the transition contributions from the quarks $u$ and $d$ inside the diquark cancel each other.

\subsection{EMT coupling vertex}\label{PionCloud}

    In Sec. \ref{QDApproach}, $\Lambda^{\mu\nu}$ is introduced as the coupling vertex between the quark and the EMT current. Assuming that the quark is point-like, the vertex can be derived from the quark Lagrangian, which can be referred to Ref. \cite{Fu:2022rkn}, and the obtained point-like vertex reads
    \begin{equation}
    \label{BareVertex}
        \Lambda^{\mu\nu}_{PL}=\frac{1}{2}\left(l^\mu \gamma^\nu+l^\nu \gamma^\mu\right),
    \end{equation}
    where $l$ is the integral variable in Eq. \eqref{QuarkT}.

However, it is assumed that the quark is no longer a point-like particle, but a so-called dressed quark surrounded by the pion cloud in this paper. The pion cloud correction is investigated based on the NJL model in Ref. \cite{Cloet:2014rja}. Similar with the diquark, the inner structure of the quark is considered and the interaction vertex between the quark and the EMT current is modified accordingly, which reads
    \begin{equation}
    \label{gqvertex}
        \Lambda^{\mu\nu} \left( t \right)=\frac{1}{2} l^{\{\mu} \gamma^{\nu\}} F_{1,0}^q(t)+\frac{\left(q^\mu q^\nu-g^{\mu\nu}q^2\right)}{m_q}F_{2,0}^q(t)+m_q g^{\mu\nu}F_{3,0}(t)+\frac{i}{4m_q}l^{ \{ \mu}\sigma^{\nu\}q}F_{4,0}^q(t),
    \end{equation}  
    where the form factors $F_{i,0}^q(t)$ are the mixture between the point-like quark and the dressed quark. Assuming that the quark discussed is on-shell, Eq.~\eqref{gqvertex} can be transformed into the same structure as Eq.~\eqref{TMatrix} with the on-shell identities in Ref.~\cite{Lorce:2017isp}. As discussed in Refs. \cite{Cloet:2014rja,Horikawa:2005dh}, the on-shell approximation only can be formulated with the pole approximation for the $t$-matrix of pion. 

    \begin{figure}[H]
	   \centering
	   \includegraphics[width=0.9\linewidth]{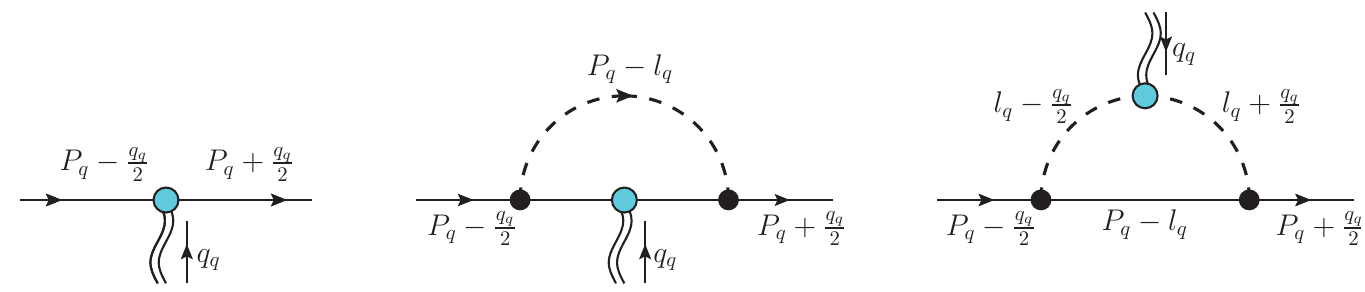}
	   \caption{\small{The coupling process between the EMT and the quark with the pion cloud correction, where $P_q=(p'_q+p_q)/2$, $q_q=p'_q-p_q=q$ and $p_q$ ($p'_q$) is the initial (final) momentum of the dressed quark.} The first panel shows the coupling process without the pion cloud. In the second and third panels, the EMT current respectively couples with the quark and the pion, and the form factors $f_{i,0}^{q/\pi}\, (i=1,2,4)$ are originated from the two panels. Specially note that when calculating the second and the third figures, the quark coupling with the pion is assumed to be point-like to simplify the calculation. The blue pionts stand for the bare vertex in Eq. \eqref{BareVertex}.  }
	   \label{pcCoupling}
    \end{figure}

\begin{figure}[H]
    \centering
    \includegraphics[width=0.3\linewidth]{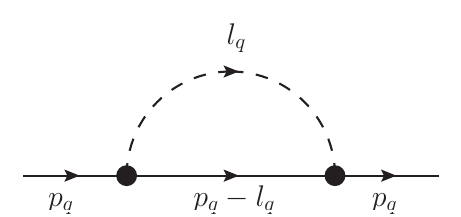}
    \caption{\small{Quark self-energy diagram. The solid line stands for the quark and the dashed curve stands for the pion. The dark dot symbolises the quark-pion vertex $\gamma_5 \tau_i$.}}
    \label{PionLoop}
    \end{figure}

    The form factors $F_{i,0}^q(t)$ can be expressed as
    \begin{equation}
    \label{gqvertex124}
        F_{i,0}^q=Z F_{i,0}^{PL} +\left(1-Z\right)(f_{i,0}^q +f_{i,0}^\pi),\quad i=1,2,3,4,
    \end{equation} 
    where $F_{1,0}^{PL}=1$, $F_{k,0}^{PL}=0$ ($k=2,3,4$) are the form factors of the point-like quark, and $f_{i,0}^q$ and $f_{i,0}^{\pi}$ ($i=1,2,3,4$) are the form factors originated from the second and third figures in Fig.~\ref{pcCoupling} respectively. The factor $Z$ stands for the possibility to strike a quark without the pion cloud~\cite{Cloet:2014rja}, and it is derived from the self-energy $\Sigma(p)$ of the pion loop as
    \begin{equation}
    \begin{split}
    \label{pcz}
        Z=1+\left.\frac{\partial \Sigma(p_q)}{\partial \slashed{p}_q}\right|_{\slashed{p}_q=m_q}.
    \end{split}
    \end{equation}  
    The self-energy is originated from Fig.~\ref{PionLoop}, which reads
    \begin{equation}
    \label{pcSelfEnergy}
        \Sigma(p_q)= - \int\frac{d^4 l_q}{(2\pi)^4}\gamma^5\tau_i S(p_q-l_q) \gamma^5\tau_i \tau_\pi(l_q),
    \end{equation}  
    where $\tau_\pi(p)$ is the NJL $t$-matrix of pion with the pole approximation as 
    \begin{equation}
    \label{taupi}
        \tau_\pi(p)=\frac{i Z_\pi}{p^2-m_\pi^2+i\varepsilon}.
    \end{equation}  
    A more detailed introduction of $\tau_\pi$ and the definition of $Z_{\pi}$ are presented in Ref.~\cite{Cloet:2014rja}.

    Similarly, for the electromagnetic current $J^\mu$, the structure of the effective vertex reads $\Lambda^\mu \left( t \right)=\gamma^\mu F_{1}^q\left( t \right)+\frac{i \sigma^{\mu q}}{2M_q} F_{2}^q\left( t \right)$. The detailed definitions of $F_1^q(t)$ and $F_2^q(t)$ can be referred to Ref.~\cite{Cloet:2014rja}.

    When calculating the quark and diquark contribution in Sec. \ref{QDApproach}, an additional scalar function (seen in Eq. \eqref{sfunc}) is attached to the vertex to simulate the bound state of the nucleon. However, since the pions are not confined, we do not use the scalar function here. During the calculation of the pion cloud correction, we choose the proper-time regularization scheme with the relation,
    \begin{equation}
    \begin{split}
    \label{propertime}
        \frac{1}{X^n}=\frac{1}{(n-1)!}\int^{\infty}_0 d \tau \tau^{n-1} e^{-\tau X}\rightarrow \frac{1}{(n-1)!}\int^{1/\Lambda_{IR}^2}_{1/\Lambda_{UV}^2} d \tau \tau^{n-1} e^{-\tau X},
    \end{split}
    \end{equation}
    where $\Lambda_{IR}$ and $\Lambda_{UV}$ are the two model-dependent 
    cut-off parameters.
    
    Different with the additional scalar function we used before, the proper-time regularization could maintain the gauge invariance and the Ward-Takahashi identity. The cut-off parameters $\Lambda_{IR}$ and $\Lambda_{UV}$ are borrowed from Ref. \cite{Cloet:2014rja}, where $\Lambda_{IR}$ takes the role to eliminate the unphysical thresholds 
    for the hadrons decay to simulate the quark confinement in QCD. 
    Since the pion and quark inside the dressed quark are not 
    confined, the cut-off parameters are determined as $\Lambda_{UV}=0.645\,\text{GeV}$ and $\Lambda_{IR}=0\,\text{GeV}$~\cite{Cloet:2014rja} when calculating 
    the second and third figures in Fig. \ref{pcCoupling}. 
    When considering the inner structure of the pion, where pion is 
    regarded as a bound state of quark-antiquark pair (similar with Fig. \ref{td0}), we choose $\Lambda_{IR}=0.240\, \text{GeV}$.

\section{Numerical Results}
\subsection{Parameter determination}
\label{ParaD}
\begin{figure}[htbp]
\centering
\includegraphics[width=0.46\linewidth]{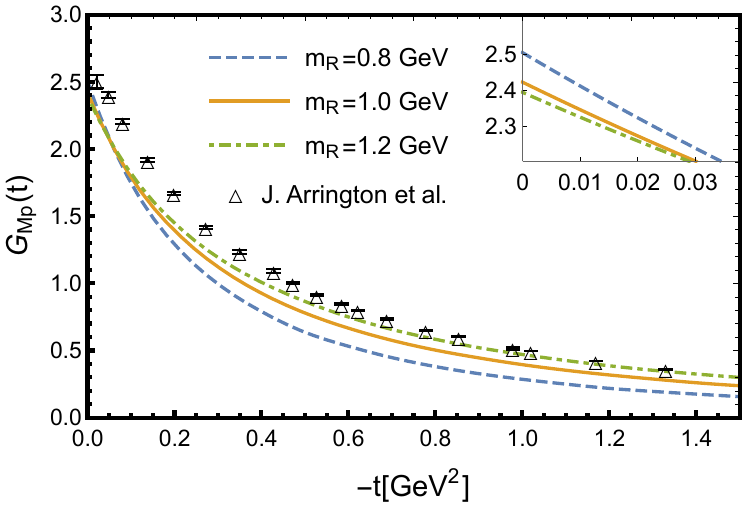}
\caption{\small{$G_M(t)$ of the proton with different $m_R$. The dashed, solid and the dotted-dashed curves respectively represent the results with $m_R=0.8$ GeV, 1.0 GeV, and 1.2 GeV. The triangles represent the data extracted from the experiment in Ref.~\cite{Arrington:2007ux}. The small figure in the upper right corner shows $G_{Mp}(t)$ when $t \approx 0$.}}
\label{mR}
\end{figure}

To ensure that the quark and the diquark exist in the bound state, the input masses must satisfy the relations $M_N<m_q+m_D$ and $m_D<2 m_{q}$. The masses of the nucleon and the pion are taken from Ref.~\cite{ParticleDataGroup:2024cfk}. Assuming that the scalar and the axialvector diquarks have the same masses and $m_u=m_d=m_q$, we choose $m_D=0.7$ GeV and $m_q=0.4$ GeV. With the mass parameters determined, the factors $Z$ and $Z_\pi$ in Sec. \ref{PionCloud} are calculated  as $Z=0.79$ and $Z_\pi=15.8$. $m_R$ is a cut-off parameter used in Eq.~\eqref{sfunc}, here $m_R$ is determined around the mass of the nucleon. Figure~\ref{mR} shows the $G_M(t)$ of the proton for values of $m_R$ ranging from 0.8 GeV to 1.2 GeV.
When $m_R$ goes up in a certain range, both the magnetic moment $\mu$ and the magnetic radius decrease.
Through fitting the experimental data in Ref.~\cite{Arrington:2007ux,Schiavilla:2001qe,JeffersonLabE95-001:2006dax}, we choose $m_R=1.0$ GeV.

With the normalization condition of the proton EMFF $G_{Ep}(0)=1$, the normalization constants $C_s=cc_s$ and $C_a=cc_a$ introduced in Eq.~\eqref{Vertex} and Eq.~\eqref{sfunc} are determined respectively with $\theta=0$ and $\theta=\pi/2$. It should be mentioned that the form factor $G_{En}(0)$ may not be normalized to zero strictly since the momentum-dependent scalar function in Eq.~\eqref{sfunc} breaks the Ward-Takahashi identity. 


The parameters $\theta$ and $g_1$ are determined through fitting the experimental data. In our results, the radii and magnetic moments provided merely by the scalar diquark are smaller than the experimental results, especially for the magnetic moments. On the contrast, the introduction of the axialvector diquark could raise the magnetic moments and the radii. With regard to the coupling $g_1$, the EMFFs are not sensitive to it, except the charge form factor of the neutron. After fully considering the electromagnetic properties including the magnetic moments, charge radii, and the normalization of $G_{E0}(0)$ of the proton and the neutron, we choose $\text{sin}^2 \theta=0.50$ and $g_1=0.80$, which means that the proportion of the scalar diquark is equal to the axialvector one.

The masses of different particles and partons and parameters used in this study are summarized and listed in Table \ref{para}.

\begin{table}[htbp]
    \renewcommand\arraystretch{1.3}
    \centering
    \caption{\small{Parameters used in this work.}}
    \begin{tabular}{ p{1.5cm}<{\centering} p{1.5cm}<{\centering} p{1.5cm}<{\centering} p{1.5cm}<{\centering} p{1.5cm}<{\centering} p{1.5cm}<{\centering} p{1.5cm}<{\centering} p{1.2cm}<{\centering} p{0.8cm}<{\centering}  }
        \toprule
        \toprule
            $M_N$/GeV & $m_D$/GeV & $m_q$/GeV & 
        $m_\pi$/GeV & $m_R$/GeV & $C_s/\text{GeV}^4$ & $C_a/\text{GeV}^4$ & $\sin^2 \theta$ & $g_1$ \\
        \midrule
        0.938 & 0.70 & 0.38 & 0.14 & 1.0 & 2.55 & 1.48 & 0.50 & 0.80 \\
        \bottomrule
        \bottomrule
    \end{tabular}
    \label{para}  
\end{table} 

\subsection{EMFFs of the nucleon}

\begin{figure}[htbp]
    \centering
    \includegraphics[width=0.46\linewidth]{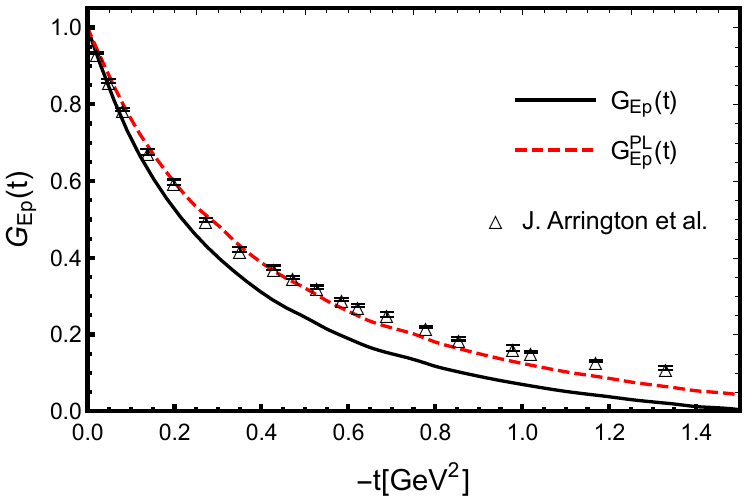}
    \includegraphics[width=0.46\linewidth]{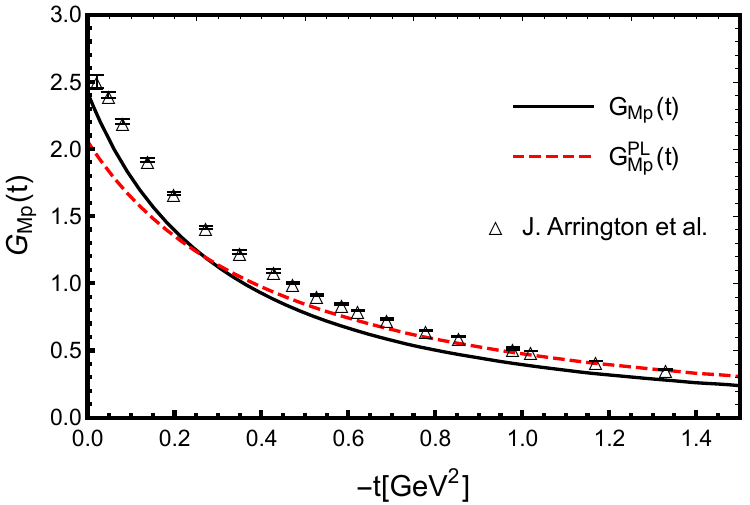}
    \includegraphics[width=0.46\linewidth]{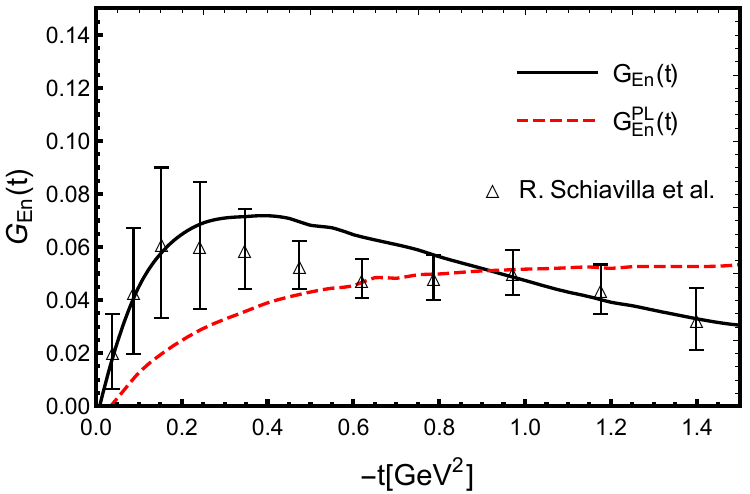}
    \includegraphics[width=0.46\linewidth]{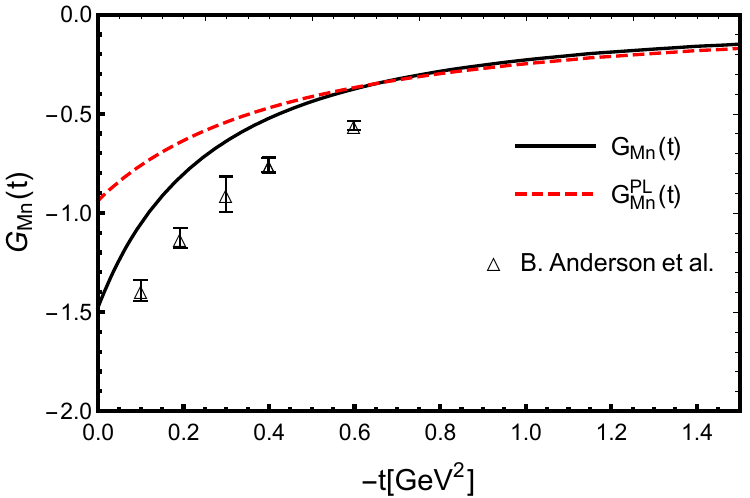}
    \caption{\small{Calculated EMFFs of the proton and the neutron. The black solid curves represent the total EMFFs calculated with the pion cloud effect, while the red dashed curves are those obtained with the point-like quark. The results are compared with the data (empty triangles) extracted from the experiments in Refs.~\cite{Arrington:2007ux,Schiavilla:2001qe,JeffersonLabE95-001:2006dax}.}}
    \label{EMFigure}
\end{figure}

The calculated EMFFs are illustrated with black solid curves in Fig.~\ref{EMFigure}, compared with the data extracted from the experiments~\cite{Arrington:2007ux,Schiavilla:2001qe,JeffersonLabE95-001:2006dax}. The magnetic moments of the nucleon are
\begin{equation}
\label{MagneticMomentResults}
    \mu_p=2.42\,\mu_N \quad \quad \mu_n=-1.48\,\mu_N,
\end{equation} 
and the charge radii derived from $G_{E}(t)$ are
\begin{equation}
\label{ChargeRadiusResults}
    \sqrt{\langle r^2_{Ep}\rangle}=0.89\,\text{fm} \quad \quad \langle r^2_{En}\rangle= -0.16\,\text{fm}^2.
\end{equation}
The detailed electromagnetic properties, including magnetic moments and electromagnetic radii, are listed in Table~\ref{EMFFsTable}. 

\begin{table}[htbp]
    \renewcommand\arraystretch{1.3}
    \centering
    \caption{\small{Electromagnetic properties of the nucleon. The first and second rows respectively list the electromagnetic properties calculated with and without the pion cloud effect.}}
    \begin{tabular}{p{2.6cm}<{\centering} p{1.4cm}<{\centering} 
    p{1.8cm}<{\centering} p{1.8cm}<{\centering} p{1.6cm}<{\centering} 
    p{1.8cm}<{\centering} p{1.8cm}<{\centering}}
        \toprule
        \toprule
         & $\mu_p /\mu_N$ & $\sqrt{\langle r^2_{Ep}\rangle}/ \text{fm}$ & $\sqrt{\langle r^2_{Mp}\rangle}/ \text{fm}$ & $\mu_n/\mu_N$ & $\langle r^2_{En}\rangle/ \text{fm}^2$ & $\sqrt{\langle r^2_{Mn}\rangle}/ \text{fm}$\\
        \midrule
            dressed quarks & 2.42 & 0.89 & 0.89 & -1.48 & -0.16 & 0.94\\
        point-like quarks & 2.06 & 0.78 & 0.76 & -0.94 & -0.03 & 0.73\\
        \midrule
        PDG~\cite{ParticleDataGroup:2024cfk} & 2.79 & 0.84 & 0.851$\pm$0.026 & -1.91 & -0.12 & $0.864^{+0.009}_{-0.008}$\\ 
        \bottomrule
        \bottomrule
    \end{tabular} 
    \label{EMFFsTable}  
\end{table}

The obtained results without the pion cloud are illustrated as the red dashed curves in Fig.~\ref{EMFigure}. The pion cloud effect increases the charge radii of both the proton and the neutron. Especially for the neutron, it distinctly changes the shape of the electric form factor, resulting in an increase in the magnitude of the charge radius and a depression of the form factor at large $\left|t\right|$. It is consistent with our intuitive expectation that the pion cloud surrounding the quark could raise the radius of the nucleon, and it is consistent with other study focusing on the pion cloud~\cite{Cloet:2014rja}. With regard to the magnetic form factors, the obtained results are smaller than the experiment results. It may be attribute to the quark-diquark model we use, which is a simple model to facilitate our calculation. Consequently, some ingredients may be ignored, such as the so-called seagull contributions \cite{Oettel:1999gc} during the scatter. Moreover, the pion cloud increases the magnitudes of the magnetic moments of the proton and the neutron by about $15\%$ and $36\%$, respectively, while the effect is insignificant at large $\left|t\right|$ to the neutron. As introduced in Sec. \ref{PionCloud}, the photon-quark vertex has two form factors $F_{1,2}^q (t)$ due to the pion cloud effect. The normalization condition ensures that $F_{1}^q (0)$ of both $u$ and $d$ quarks are equal to their charges, while $F_{2}^q (0)$ of $d$ is relatively larger than that of $u$. Therefore, compared to the proton, the pion cloud leaves more significant influence on the neutron.

\subsection{GFFs of the nucleon}

\begin{figure}[htbp]]
    \centering
    \includegraphics[width=0.46\linewidth]{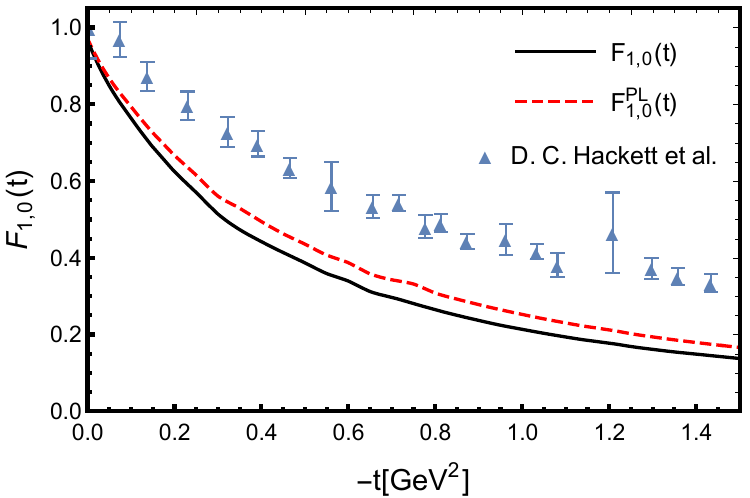}
    \includegraphics[width=0.46\linewidth]{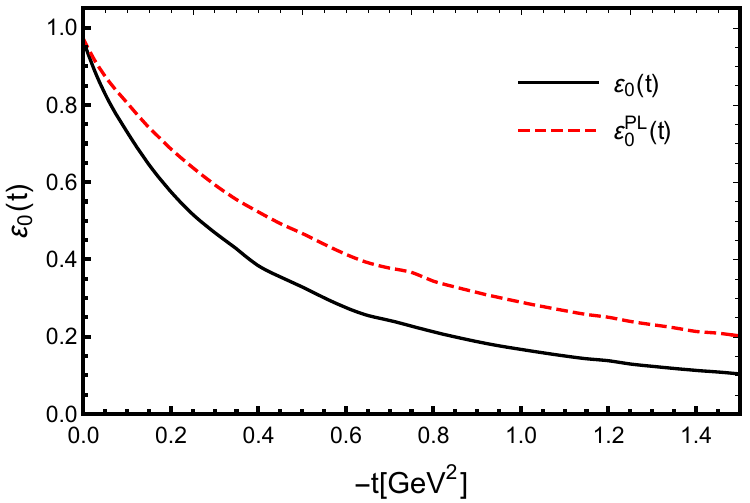}        
    \includegraphics[width=0.46\linewidth]{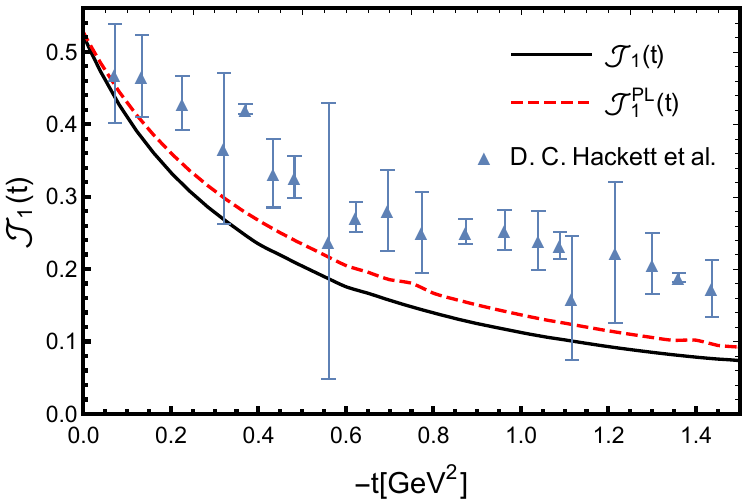}
    \includegraphics[width=0.47\linewidth]{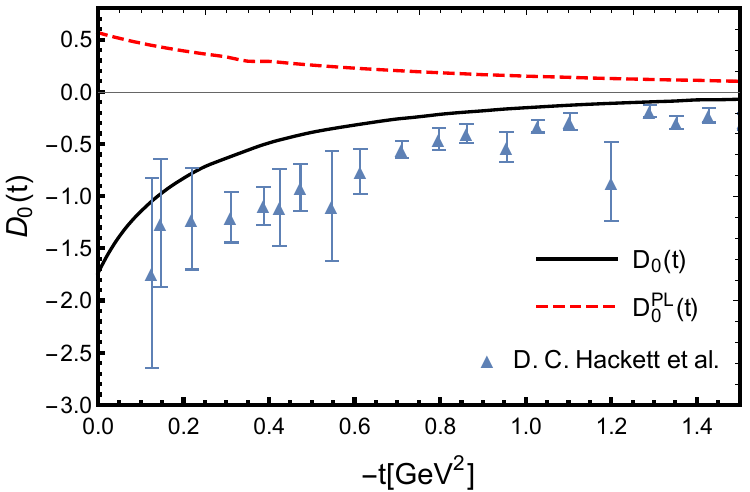}
    \caption{\small{Calculated GFFs of the nucleon, including $F_{1,0}(t)$ in Eq. \eqref{TMatrix}, $\varepsilon_0(t)$, $\mathcal{J}_1(t)$ and $D_0(t)$. Specially note that $F_{1,0}(t)$ is additionally illustrated to be compared with the lattice result. The black solid curves represent the total GFFs calculated with the pion cloud effect, while the red dashed curves are those obtained with the point-like quark. The results are compared with the lattice data (blue triangles) in Ref.~\cite{Hackett:2023rif}, which is contributed by the quarks and gluons, obtained with pion mass $m_\pi=170\,\text{MeV}$.}}
    \label{GFigure}
\end{figure}

The numerical results of the GFFs are illustrated in Fig.~\ref{GFigure}. The GFFs are independent of the charge, and thus those of the proton and the neutron are indistinguishable. The normalization constants employed here are the same as those used during the calculation of the EMFFs to maintain the consistency. However, the momentum-dependent scalar function introduced in Eq.~\eqref{sfunc} may break the gauge invariance and the Ward-Takahashi identity~\cite{Broniowski:2008hx,Davidson:1994uv}. As a result, the form factors $G_E(t)$ and $\varepsilon_0(t)$ cannot be normalized at the same time. It is seen that $\varepsilon_0(0)\approx 0.97$ and $\mathcal{J}_1(0)\approx 0.52$, which are in close agreement with the normalization condition $\varepsilon_0(0)=1$ and $\mathcal{J}_1(0)=1/2$.
Moreover, the mass and mechanical radii compared to the different models are listed in Table~\ref{MassRadiusTable} . Compared to the electromagnetic radius of the proton in Table~\ref{EMFFsTable}, the mass radius is slightly larger, which is consistent with Ref.~\cite{Cao:2024zlf}. The GFFs obtained without the pion cloud effect are illustrated with the red curves in Fig.~\ref{GFigure}, which 
still nearly satisfy the normalization condition. Similar to the charge radius, the pion cloud raises the mass radius of the nucleon by about $17\%$. 

\begin{table}[htbp]
    \renewcommand\arraystretch{1.2}
    \centering
    \caption{\small{Calculated mass and mechanical radii of the nucleon, compared with that evaluated without the pion cloud and the results from the lattice QCD~\cite{Hackett:2023rif}, the chiral quark-soliton model~\cite{Goeke:2007fp}, the Skyrme model~\cite{Kim:2012ts}, the light-front quark-diquark model~\cite{Chakrabarti:2020kdc,Choudhary:2022den}, and the $\pi-\rho-\omega$ soliton model~\cite{Jung:2014jja}.}} 
    \begin{tabular}{p{1.8cm}<{\centering} p{1.8cm}<{\centering} p{1.8cm}<{\centering} 
    p{1.8cm}<{\centering} p{1.3cm}<{\centering} p{1.3cm}<{\centering} p{1.3cm}<{\centering} p{1.3cm}<{\centering} p{1.3cm}<{\centering}}
        \toprule
        \toprule
         & dressed & point-like & \cite{Hackett:2023rif} & \cite{Goeke:2007fp} & \cite{Kim:2012ts}   &\cite{Chakrabarti:2020kdc} & \cite{Choudhary:2022den}&\cite{Jung:2014jja}\\
        \midrule
        $\sqrt{\langle r^2_m \rangle}/\text{fm}$ & 0.92 & 0.76 & $0.74 \pm 0.04$ & 0.82 & 0.82  & $\cdots$ & 0.57 & 0.88\\ 
        \midrule
        $\sqrt{\langle r^2_{\text{mech}} \rangle}/\text{fm}$ & 0.91 & 0.40 & $0.75 \pm 0.08$  & $\cdots$  & $\cdots$ & 0.86 & 0.50& $\cdots$\\ 
        \bottomrule
        \bottomrule
    \end{tabular}
    \label{MassRadiusTable}  
\end{table}

$D_{0}(t)$ is believed to be connected with the pressure and shear force in the classical physical concept~\cite{Polyakov:2018zvc}. In Fig.~\ref{GFigure}, the obtained D-term is $D=-1.74$ with the pion cloud effect. However, when only considering the point-like quark contributions, the D-term is positive. Both pressures obtained through the negative and positive D-terms in this study could satisfy the von Laue condition, but as Ref.~\cite{Perevalova:2016dln} discussed, the D-term is argued to be negative to maintain the mechanical stability of the system. This sign problem also occurs in our previous studies on the decuplet baryons~\cite{Fu:2023ijy,Wang:2023bjp}, where the meson cloud was not 
taken into account. In our previous work \cite{Fu:2023ijy}, we argue that the definitions of the pressure and shear force are only applicable in the multibody systems, while in the constituent quark model we are dealing with, the nucleon is assumed to be a few-body system composed of a quark and a diquark. However, the pion cloud, working as a gas-like composition surrounding the quark, may change the structure of our model and give the pressure and shear force clear physical meanings. The mechanical radius derived from $D_0(t)$ is compared with those in other studies in Table \ref{MassRadiusTable}. The mechanical radius is slight smaller than the mass radius, and similar with the mass and electromagnetic radii, the dressed mechanical radius is larger than the point-like one.

In the $r$-space, distributions of the mechanical properties inside the nucleon can be derived through the Fourier transformation. To consider the local particle, an additional wave packet is attached to the GFFs. Refs.~\cite{Epelbaum:2022fjc,Diehl:2002he,Freese:2021mzg} suggest that the local density distribution must depend on the size of the wave packet of the system. Furthermore, an additional wave packet is necessary physically and mathematically to guarantee the convergence of the Fourier transformation. Here we simply employ a Gaussian-like wave packet $e^{t/\lambda^2}$~\cite{Ishikawa:2017iym}, where $\lambda$ has the mass dimension and $1/\lambda$ correlates with the size of the hadron. As Ref. \cite{Burkert:2018bqq} discussed, the positive and negative pressure are separated by a zero crossing point $r_0$. In this work, we use the mass radius to characterize the crossing point, and we choose $\lambda=0.85$ GeV to ensure $r_0 \sim \sqrt{\langle r^2_m \rangle}$.

\begin{figure}[htbp]
    \centering
    \subfigure[]{\includegraphics[width=0.46\linewidth]{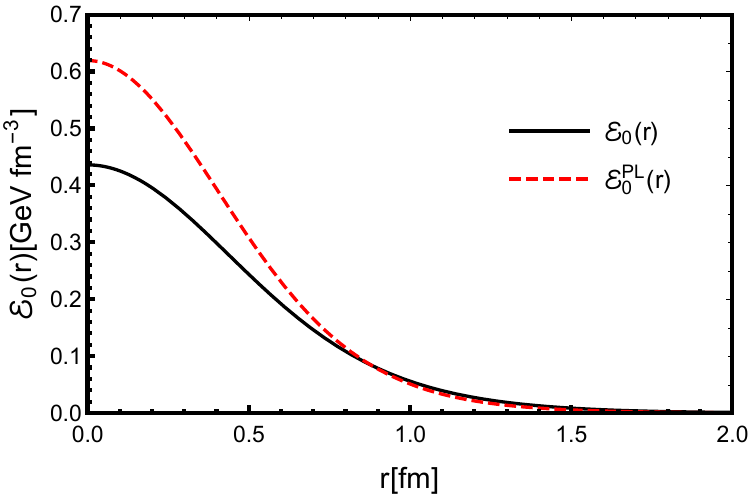}\label{ep0r}}
    \subfigure[]{\includegraphics[width=0.47\linewidth]{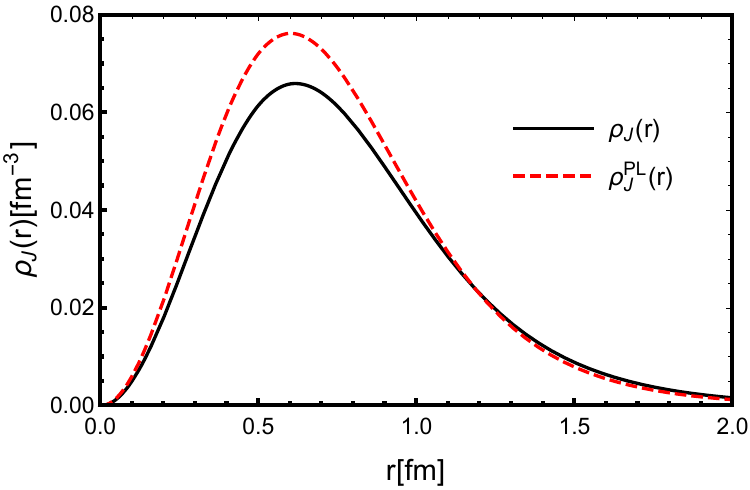}\label{rhoj}}
    \caption{\small{Panels (a) and (b) present the energy and angular momentum densities of the nucleon, where the black solid and red dashed curves respectively represent the densities with and without the pion cloud effect.}}
    \label{EAMDens}
\end{figure}

Figure~\ref{EAMDens} shows the energy densities and the angular momentum densities in the $r$-space. Comparing the two curves in Figs. \ref{ep0r} and \ref{rhoj}, the pion cloud depresses the densities in small $r$, while in larger $r$ (approximately when $r > 1\, \text{fm}$), the densities with the pion cloud effect are slightly larger. The results indicate that the pion cloud may distract the mass and angular momentum distribution from the origin. As illustrated in Fig.~\ref{GFigure}, the pion cloud surrounding the quark raises the mass and angular momentum radii, thereby shifting their distribution farther away, which is consistent with our intuitive expectations.

\begin{figure}[htbp]
    \centering
    \subfigure[]{\includegraphics[width=0.46\linewidth]{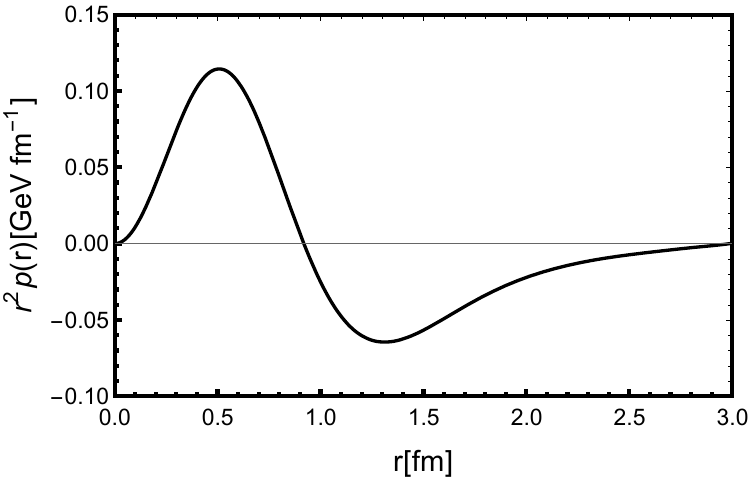}\label{PDistirbution}}
    \subfigure[]{\includegraphics[width=0.45\linewidth]{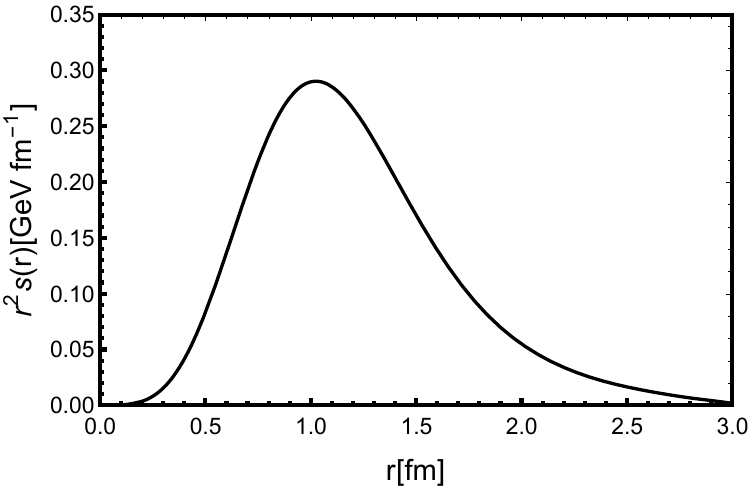}\label{SFDistirbution}}
    \caption{\small{The pressure and shear force distributions inside the nucleon, with the pion cloud effect.}}
    \label{PSF}
\end{figure}

The pressure and shear force distributions introduced in Eqs. \eqref{Pressure} and \eqref{ShearForce} are shown in Fig.~\ref{PSF}.
As seen in the black curve in Fig. \ref{PDistirbution}, the pressure distribution is 
quantatively consistent with that presented in Ref. \cite{Burkert:2018bqq}. The repulsive (positive) and binding (negative) pressures are separated by the zero crossing point, and as discussed before, the crossing point is approximately equal to the mass radius $\sqrt{\langle r^2_m \rangle}=0.92 \, \text{fm}$. Moreover, the peak and the valley of the pressure are $r=0.51\, \text{fm}$ and $r=1.31\, \text{fm}$. In our previous work \cite{Fu:2023ijy}, we argue that the so-called pressure and shear force obtained with the point-like quark may not have clear physical meanings in the constituent quark model. Therefore, they are not illustrated in Fig.~\ref{PSF}. As the same as the D-term, the pion cloud change the sign of the pressure and the shear force inside the nucleon. Similar with other mechanical properties such as the angular momentum density, the pion cloud effect drives the maximum and the minimum of the pressure and shear force distributions farther from the origin. 

\section{Summary And Discussion}
In this work, both the EMFFs and GFFs of the nucleon have been simultaneously calculated with a relativistic covariant quark-diquark approach, which could simplify the nucleon structure from a three-body system into a two-body system. An additional scalar function is employed to simulate the bound state between the quark and the diquark instead of solving the Bethe-Salpeter equation directly. Furthermore, the pion cloud effect is taken into consideration to obtain more reasonable results. In this work, it is assumed that the quark is no longer a point-like particle, but a so-called dressed quark coupling with the pion. Considering the quark internal structure, the coupling vertex between the quark and the electromagnetic or the EMT current is modified.

The parameters used in the paper are determined through fitting the experimental data of the EMFFs. The obtained EMFFs results are found to be in reasonable agreement with the experimental data, except that the magnetic moments are smaller, which may attribute to the simple quark-diquark model. The pion cloud plays an important role, increasing the magnetic form factors of both the proton and the neutron significantly. Moreover, the magnitude of the electromagnetic radii, particularly for the neutron, is raised.

Regarding the GFFs, the pion cloud leaves a significant influence on the D-term, even changing its sign from positive to negative. We believe that the classical definitions of the pressure and shear force may not be applicable in the quark-diquark approach adopted in the few-body system. However, the pion cloud surrounding the quarks may transform the nucleon into a more complex system, giving the pressure and shear force clear physical meanings. Therefore, the introduction of the pion cloud is indispensable in our model. With the pion cloud correction, the mass radius increases about $17\%$, still smaller than the charge radius of the proton. Furthermore, the mechanical properties including the energy density, the angular momentum density, the mechanical radius, and the pressure and shear force distributions, are also given in the coordinate space by the Fourier transformed form factors.

It should be mentioned that in our study, the pion cloud effect is evaluated with a simple model. Although the nucleon are composed of the dressed quarks, the EMT-quark vertex used in the second and third panels of Fig.~\ref{pcCoupling} is still the piont-like vertex, since we assume that the pion cloud corrections result from the higher order terms is much weaker. A more complete interaction vertex can be referred in Ref. \cite{Freese:2019bhb}, which focuses on the GFFs of the light mesons and gives a vertex with all the quarks dressed through solving the BSE. Therefore, if fully considering the pion cloud effect on all the quarks involved in the future study, more reasonable results may be obtained.

With the nucleon form factors obtained, our future studies will focus on the $N-\Delta$ transition form factors with the pion cloud effect. In our previous papers~\cite{Wang:2023bjp,Fu:2022rkn} about the form factors of $\Delta$, the pion cloud effect was not considered. It is expected that the introduction of the pion cloud could provide more reasonable information about the properties of $\Delta$ and the $N-\Delta$ transition.

\section*{Acknowledgements}
We are grateful to Ian C. Cloët for the constructive discussions. This work is supported by the National Key Research and Development Program of China under Contracts No. 2020YFA0406300, the National Natural Science Foundation of China under Grants No. 11975245, No. 12375142 and No. 12447121 and the Gansu Province Postdoctor Foundation.

\bibliographystyle{unsrt}
\bibliography{references}

\end{document}